\documentclass[acmsmall,nonacm]{acmart}

\AtBeginDocument{%
  \providecommand\BibTeX{{%
    Bib\TeX}}}

\setcopyright{none}
\copyrightyear{2026}
\acmYear{2026}
\acmDOI{XXXXXXX.XXXXXXX}
\acmConference[ACM Conference '26]{Conference}{December 2026}{}

\usepackage{amsmath}
\usepackage{algorithmic}
\usepackage{graphicx}
\usepackage{textcomp}
\usepackage{xcolor}
\usepackage{subcaption} 
\usepackage{wrapfig,lipsum,booktabs}
\usepackage{adjustbox}

\usepackage[normalem]{ulem}
\def\BibTeX{{\rm B\kern-.05em{\sc i\kern-.025em b}\kern-.08em
    T\kern-.1667em\lower.7ex\hbox{E}\kern-.125emX}}

\newcommand{\ie}{\textit{i.e.,}\ }
\newcommand{\eg}{\textit{e.g.,}\ }
\newcommand{\myverbS}{\fontsize{8}{47}\usefont{OT1}{lmtt}{b}{n}\noindent }
\newcommand{\myverb}{\fontsize{9}{48}\usefont{OT1}{lmtt}{b}{n}\noindent }
\begin{document}
\title[Semantic Matching of Behavioral Primitives for MUD-Based IoT Device Identification]{Semantic Matching of Behavioral Primitives for MUD-Based IoT Device Identification}

\author{Samuel Witt}
\affiliation{%
	\institution{School of EE\&T, UNSW Sydney}
	\city{Sydney}
	\country{Australia}
}
\email{s.witt@student.unsw.edu.au}

\author{Hassan Habibi Gharakheili}
\affiliation{%
	\institution{School of EE\&T, UNSW Sydney}
	\city{Sydney}
	\country{Australia}
}
\email{h.habibi@unsw.edu.au}

\renewcommand{\shortauthors}{S. Witt and H. Habibi Gharakheili}

\begin{abstract}
Accurate identification of Internet of Things (IoT) devices, such as cameras, lightbulbs, and voice assistants, is important for security management and policy enforcement. Existing approaches identify devices from packet- or flow-level traffic characteristics. Manufacturer Usage Description (MUD) profiles provide a complementary, policy-level representation of intended device communication via Access Control Entries (ACEs), which encode protocol, endpoint, direction, and port semantics. While exact ACE matching is effective when observed behavior overlaps with the reference profile, it provides little evidence when that overlap becomes sparse.
This paper investigates the semantic matching of individual ACEs as behavioral primitives for MUD-based IoT device identification. Our specific contributions are threefold. 
(1) Using 28 publicly available MUD profiles containing 1,023 ACE instances, we show that representing individual ACEs preserves behavioral distinctions that are obscured when an entire profile is represented by a single vector. 
(2) Through controlled experiments with unseen ACEs, endpoint perturbation, and partial observation, we characterize how semantic matching behaves as exact overlap decreases.
(3) Using real IoT traffic traces comprising more than 800,000 flows, we show that semantic matching provides stronger evidence during early and sparse-overlap observations, whereas exact matching becomes stronger as stable overlap accumulates.
These results demonstrate that semantic ACE matching can complement exact MUD matching when runtime observations provide limited literal overlap with reference profiles.
We release the code and research artifacts to support reproducibility.
\end{abstract}

\ccsdesc[500]{Networks~Network management}
\ccsdesc[300]{Networks~Network measurement}
\ccsdesc[100]{Networks~Network monitoring}
\ccsdesc[100]{Security and privacy~Network security}
\keywords{IoT device identification, Manufacturer Usage Description (MUD), semantic matching, runtime traffic, network management}

\maketitle

\section{Introduction}

IoT devices are increasingly deployed in enterprise, industrial, and home networks, creating challenges for network management and security~\cite{Usenix2020iot,IoTFinder2020euroSP,ndss2025}. Accurately identifying devices connected to a  network is important for policy enforcement, access control, and anomaly detection~\cite{Lumos2022Usenix}. 
Existing identification approaches commonly infer device identity from observed network traffic using statistical packet or flow features, endpoint information, protocol sequences, or learned traffic representations~\cite{25tnse,haystack2020IMC,IPbased2018,ndss2025,22ipfixJIoT}. 
Manufacturer Usage Description (MUD) profiles~\cite{rfc8520} provide a complementary view of device behavior. Rather than describing fine-grained traffic dynamics, a MUD profile specifies intended communication behavior through Access Control Entries (ACEs), where each ACE captures protocol, endpoint, direction, and port semantics. 
MUD therefore provides a standardized, policy-level representation against which observed device communication can be compared~\cite{ACSAC2022,tdsc2022,MUDfirewall2024}.

Existing MUD-based monitoring approaches primarily assess whether observed communication is consistent with a device profile through exact policy or ACE overlap~\cite{tdsc2022,EuroSP2024}. Exact overlap provides strong identification evidence when observed behaviors match the reference profile, with confidence increasing as additional matching ACEs are observed. However, exact matching is inherently binary. In practice, runtime observations may contain only a subset of the behaviors described by a profile, and related communications may differ lexically, for example because they refer to different but related service endpoints or hostnames~\cite{EuroSP2024}. 
For example, two communications may use the same transport protocol, service port, direction, and related cloud service while referring to different hostnames. Exact matching assigns no similarity to such observations despite their shared behavioral characteristics. We therefore ask: \textit{whether the semantics encoded by individual communication rules can still provide useful evidence about device identity when literal ACE overlap is limited.}

\begin{wrapfigure}{r}{0.60\textwidth}
	\centering	{\includegraphics[width=0.95\linewidth]{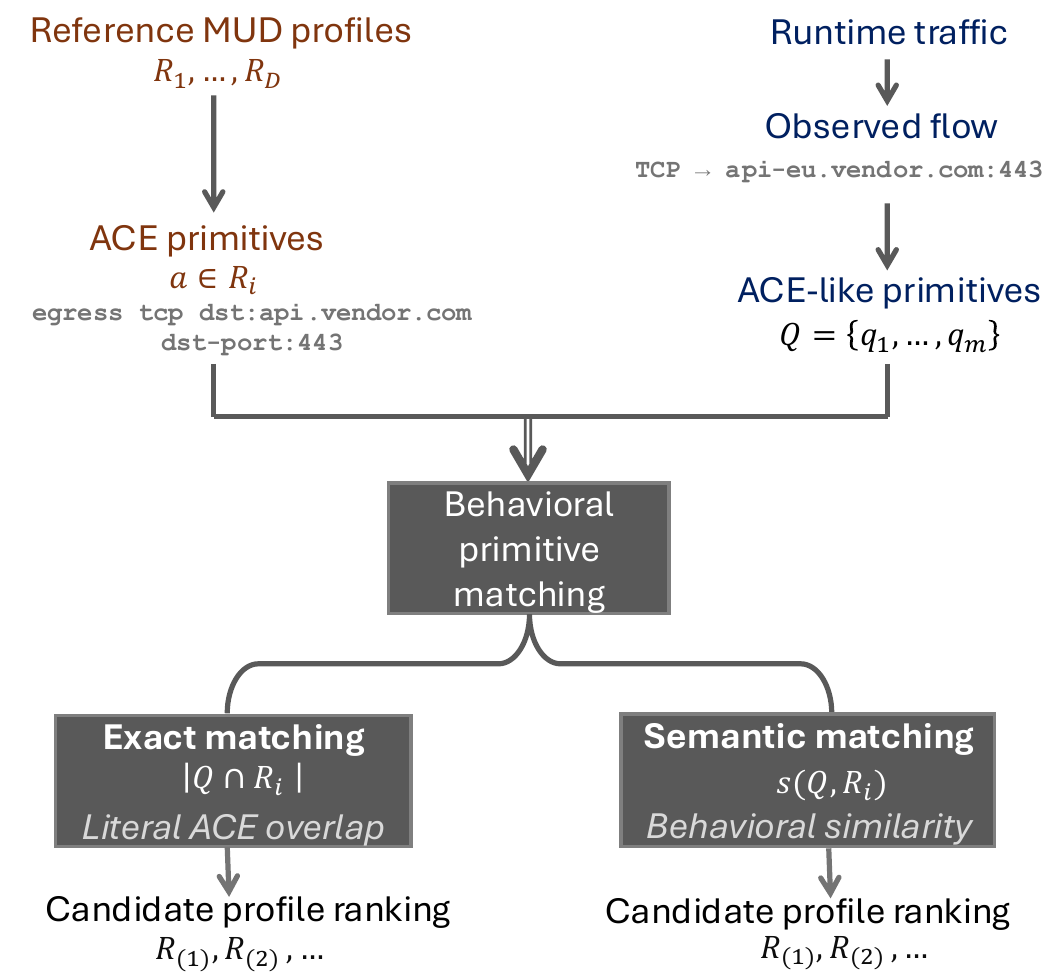}}
	\vspace{-2mm}
	\caption{Overview of the behavioral primitive matching framework. Reference MUD profiles $R_1,...,R_D$ are represented by individual ACEs, while observed runtime flows are converted into ACE-like query primitives $Q$. We evaluate two matching strategies independently: exact matching based on literal ACE overlap and semantic matching based on behavioral similarity.}
	\vspace{-4mm}
	\label{fig:overview}
\end{wrapfigure}

We investigate this question by using individual ACEs as behavioral primitives for semantic matching. Fig.~\ref{fig:overview} illustrates the framework. 
Rather than collapsing an entire MUD profile into a single representation, we retain these primitives separately and make lexically different primitives comparable through semantic embeddings. At runtime, observed flows are converted into ACE-like behavioral primitives and compared at the ACE level against candidate reference MUD profiles, allowing related behaviors to contribute identification evidence even when they do not match exactly. Our objective is not to replace packet- or flow-based classifiers, but to investigate semantic matching as a complementary identification signal in environments where reference MUD profiles are available. This setting assumes that the network operator has access to a repository of candidate MUD profiles. Although manufacturer adoption of MUD remains limited, reference profiles can also be generated from observed device traffic using profiling tools~\cite{tdsc2022}; the profiles evaluated in this work were generated in this manner.

Our specific contributions are threefold. Our \textbf{first} contribution formulates MUD-based device identification using individual ACEs as behavioral primitives and compares their representation with alternatives that represent the whole profile as a single vector. The results show that retaining individual ACEs preserves variation among communication rules that is lost through whole-profile representation and aggregation (\S\ref{sec:representation}).
Our \textbf{second} contribution systematically characterizes exact and semantic matching under controlled changes in ACE overlap, including unseen ACEs, endpoint perturbation, and mixed partial observation. The results identify when semantic similarity retains identification evidence as literal overlap becomes sparse or disappears (\S\ref{sec:device_identification}). 
These experiments isolate the effect of decreasing ACE overlap rather than model specific operational causes such as firmware updates. 
Our \textbf{third} contribution evaluates these matching approaches on real IoT traffic traces comprising more than 800,000 flows. The results show that semantic matching of individual ACEs provides stronger identification during early observations and within short traffic windows when exact overlap is limited, while exact matching strengthens as repeated literal evidence accumulates (\S\ref{sec:runtime_flows}).
To support reproducibility and further research, we publicly release \cite{gitHub2026} the datasets, derived behavioral representations, embeddings, and implementation code associated with our evaluation. 


\section{Representing MUD Behavior for Semantic Matching}
\label{sec:representation}

We first consider the appropriate granularity for representing MUD behavior in semantic comparison. 
Representing an entire MUD profile with a single vector may obscure distinctions between individual communication behaviors, whereas preserving its constituent ACEs retains the behavioral units that can also be reconstructed from runtime traffic. We therefore examine whether individual ACEs provide a suitable granularity for semantic matching.


\subsection{Behavioral Primitives}
A MUD profile~\cite{rfc8520} is a JSON document that describes the intended network communication of a device type. 
Within a MUD profile, access control lists contain individual ACEs, each specifying a communication rule, such as a TCP connection from the device to a named cloud endpoint on a particular port.

\begin{wraptable}{r}{0.65\columnwidth}
	\centering
	\vspace{-6mm}
	\caption{Reference MUD profiles used in our evaluation.}
	\label{tab:mud_profiles}
	\vspace{-2mm}
	\small
	\resizebox{\linewidth}{!}{%
		\begin{tabular}{l r l r}
			\toprule
			\textbf{Device} & \textbf{\# ACEs} & \textbf{Device} & \textbf{\# ACEs} \\
			\midrule
			Hue lightbulb                & 38  & Netatmo camera          & 48  \\
			Netatmo weather station & 9   & SmartThings              & 14  \\
			Amazon Echo             & 57  & August doorbell camera   & 65  \\
			Awair air quality       & 15  & Belkin camera            & 43  \\
			Blipcare BP meter       & 9   & Canary camera            & 29  \\
			Chromecast Ultra        & 201 & Dropcam                  & 19  \\
			Hello barbie            & 13  & HP printer                & 42  \\
			iHome power plug        & 17  & LIFX lightbulb                 & 18  \\
			Nest smoke sensor       & 104 & Pix-Star photoframe      & 17  \\
			Ring doorbell           & 17  & Samsung smart-cam          & 37  \\
			TP-Link camera          & 33  & TP-Link plug               & 32  \\
			Triby speaker           & 43  & WeMo motion                & 31  \\
			WeMo switch             & 33  & Withings baby monitor     & 19  \\
			Withings cardio         & 9   & Withings sleep sensor      & 11  \\
			\midrule
			\textbf{Total}          & \textbf{1,023} & & \\
			\bottomrule
		\end{tabular}
	}
	\vspace{-4mm}
\end{wraptable}

We use a public dataset of MUD profiles for 28 connected devices, generated with MUDgee~\cite{tdsc2022}. The profiles span cameras, sensors, smart plugs, hubs, speakers, and other connected devices, and collectively contain 1,023 ACE instances, of which 710 are unique. As shown in Table~\ref{tab:mud_profiles}, the number of ACEs varies across profiles, ranging from 9 to 201, with a median of 30. Repeated ACEs commonly correspond to shared infrastructure behaviors, including DNS (UDP/53), NTP (UDP/123), DHCP,
and gateway communication.

MUD JSON documents contain substantial repetition in their syntax, in addition to their communication rules. Directly embedding these documents may cause the resulting representation to reflect repeated JSON syntax in addition to the communication behavior of interest. We therefore extract each ACE and express it as compact ACE text that retains its protocol, endpoint, direction, and port semantics. For example, an ACE may be represented as:

\smallskip
\noindent
``{\myverbS{egress ipv4 tcp (direction-initiated:from-device) dst:tech.carematix.com dst-port:8777}}''
\smallskip

We use BGE-M3~\cite{chen2024bgem3} to encode these representations. BGE-M3 produces 1,024-dimensional embeddings and supports semantic comparison of structured text. Converting the raw MUD JSON profiles into compact ACE text reduces the average BGE-M3 token count per profile from 4,852 to 933 (83\%), while retaining the protocol, endpoint, direction, and port fields used for behavioral matching.
We compare three representation granularities: (i) raw whole-profile JSON, (ii) compact whole-profile ACE text, and (iii) individual ACE representations.




\subsection{Representation Granularity}
\label{sec:geometry}

We next examine how representation granularity affects the separation of MUD behavior in the embedding space. This analysis does not evaluate device identification directly. Instead, it examines whether preserving individual ACEs exposes behavioral variation that is suppressed when an entire MUD profile is represented by a single vector.

\begin{wraptable}{r}{0.65\columnwidth}
	\centering
	\caption{Embedding geometry across three representation granularities for the 28 evaluated MUD profiles.}
	\vspace{-4mm}
	\resizebox{\linewidth}{!}{%
		\begin{tabular}{llll}
			\toprule
			\textbf{Metric} & \textbf{Raw JSON} & \textbf{Compact ACE text} & \textbf{Individual ACEs} \\
			\midrule
			Mean pairwise cosine & 0.936 & 0.865 & 0.698 \\
			Effective rank       & 23.3  & 24.2  & 197.6 \\
			Dims for 90\% var.   & 17    & 18    & 45 \\
			\bottomrule
		\end{tabular}
	}
	\label{tab:BGEembedding_quality}
\end{wraptable}

In Table~\ref{tab:BGEembedding_quality}, we compare the three representations introduced above using three metrics. We measure embedding concentration using mean of pairwise cosine similarity. Values close to one indicate that the vectors point in similar directions and are therefore less separable. Effective rank estimates the number of independent directions represented by the embedding matrix. We also report the number of principal components required to explain 90\% of the variance.


Raw whole-profile JSON representations are highly concentrated, with a mean pairwise cosine of 0.936. 
Converting each profile into compact ACE text reduces this value to 0.865, indicating greater separation after repeated JSON syntax is removed.
Preserving ACEs individually produces a substantially greater variation in the embedding space, reducing the mean pairwise cosine to 0.698. The effective rank increases from 24.2 for compact whole-file representations to 197.6 for individual ACEs, while the number of dimensions required to explain 90\% of the variance increases from 18 to 45.
The whole-profile matrices contain only 28 device vectors and therefore have a mean-centered rank of at most 27; ACE-level representations instead expose variation across the individual communication rules contained within those profiles. 
These results do not imply that greater embedding rank alone improves device identification. Instead, they show that representing ACEs individually preserves distinctions that are lost when the whole profile is collapsed into one vector.

\begin{wrapfigure}{r}{0.55\textwidth}
	\centering
	{\includegraphics[width=0.95\linewidth]{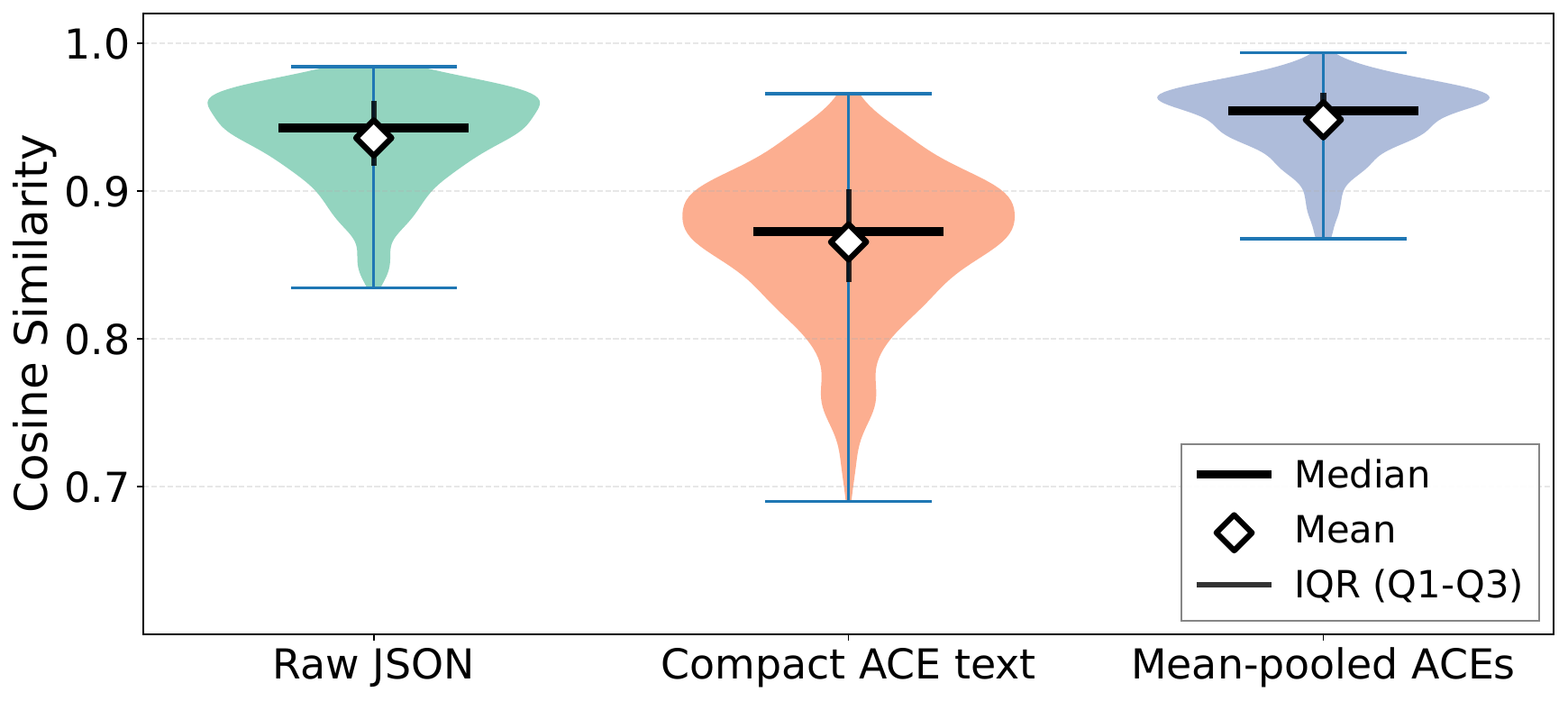}}
	\vspace{-4mm}
	\caption{Pairwise cosine similarity among device-level representations. Raw JSON and compact ACE text produce one embedding per MUD profile; mean-pooled ACEs are obtained by separately embedding individual ACEs and averaging their vectors into a device-level representation.}
	\vspace{-4mm}
	\label{fig:violin_cross_sim}
\end{wrapfigure}

Device identification ultimately requires comparing the collective behavior of an observed device against a candidate profile. A natural device-level representation is therefore to embed ACEs separately and mean-pool their vectors. Fig.~\ref{fig:violin_cross_sim} compares pairwise cosine similarity across raw whole-profile JSON, compact whole-profile ACE text, and mean-pooled ACE representations.

Much of this separation disappears after the ACE embeddings are averaged, and the resulting device representations again become highly concentrated. 
This concentration is consistent with the influence of behaviors shared across many profiles, such as DNS, NTP, DHCP, and gateway communication, whose contributions are combined with more device-specific ACEs during aggregation. Mean pooling, therefore, provides a useful baseline, but the loss of separation also motivates keeping individual ACEs during matching.
We observe similar qualitative trends using OpenAI text-embedding-3-large, providing evidence that the observed representation effects are not specific to BGE-M3. We use BGE-M3 for the remaining experiments and provide the embeddings from both encoders in our public repository \cite{gitHub2026}.

\subsection{Embedding Calibration}
\label{sec:whitening}

Semantic embeddings can exhibit anisotropy, where a small number of dominant directions cause different representations to occupy a narrow region of the embedding space~\cite{li2020bertflow,su2021whitening}. This concentration can reduce the discriminative range of cosine similarity. We therefore apply whitening decorrelation~\cite{su2021whitening} before downstream semantic matching.

Whitening parameters are estimated only from the reference ACE corpus to avoid information from evaluation queries or runtime observations from entering the calibration. The transformation recenters the embeddings, projects them onto principal components, and rescales the retained components according to their variance. We retain 256 principal components and $L_2$-normalize the resulting vectors; the complete transformation is given in Appendix~\ref{apx:sec:white}. 

Whitening reduces cosine concentration without adding behavioral information. 
For example, mean-pooled ACE signatures have a mean pairwise cosine of 0.948 and an effective rank of 23.1 before whitening. Applying whitening before pooling reduces the mean cosine to 0.073 and increases the effective rank to 26.5, close to the maximum supported by the 28 device-level signatures. All subsequent semantic matching experiments, therefore, use whitened ACE embeddings.



\section{Controlled Evaluation under Degraded ACE Overlap}\label{sec:device_identification}

The representation analysis in \S\ref{sec:representation} motivates retaining ACEs individually for semantic comparison. We now examine how exact and semantic matching behave as literal overlap between an observed ACE set and candidate MUD profiles is systematically reduced. 

Unlike the real-traffic evaluation in \S\ref{sec:runtime_flows}, the experiments in this section construct controlled queries from the reference MUD corpus. 
Their purpose is to isolate the effect of reducing ACE overlap in different ways, rather than to reproduce particular operational causes of behavioral change, such as firmware updates or software evolution.

\subsection{Controlled Query Construction}
\label{sec:runtime_settings}
Each evaluation episode consists of a query $Q$, a repository of candidate MUD profiles $\{R_1,\ldots,R_D\}$, and the known source device $d$ from which the query is constructed. The query contains one or more ACE behavioral primitives presented to the matcher. Each method scores $Q$ against every candidate profile $R_i$ and ranks the profiles accordingly. Identification is correct when the profile corresponding to $d$ is ranked first.
We construct queries under three controlled conditions that manipulate literal ACE overlap in different ways: 

\paragraph{Unseen ACEs.} This condition examines whether the query still provides evidence about the sources device when its ACEs have no exact counterpart in the candidate repository. For each episode, the selected query ACEs are removed from all candidate profiles before scoring, guaranteeing zero literal overlap for those behaviors. This leave-out construction isolates the zero-overlap condition rather than modeling changes to a deployed MUD repository. We consider a single unseen ACE, multiple unseen ACEs belonging to the same semantic family, and unseen ACEs drawn from distinct families. Appendix~\ref{apdx:unseen} describes the family construction and sampling procedure.

\paragraph{Endpoint perturbation.} This condition removes exact ACE equality by modifying domain names while preserving protocol, direction, and port semantics. Perturbations include numeric modification, regional token insertion, and domain-suffix substitution. Generated ACEs that reproduce any ACE in the reference corpus are discarded, ensuring zero exact overlap with the candidate repository. Appendix~\ref{apdx:perturb} provides the perturbation and sampling details.

\paragraph{Mixed partial observation.} This condition combines exact and non-exact evidence within the same query. For each source profile, we first sample a fraction $r \in \{0.10, 0.25, 0.50\}$ of its ACEs to form the query and perturb a fraction $p \in \{0.00, 0.25, 0.50, 1.00\}$ of the sampled domain-name ACEs. Depending on the configuration, one sampled ACE is additionally made unseen by removing its original text from the source device's reference profile for that query. The resulting queries contain different combinations of exact, endpoint-perturbed, and unseen ACEs, allowing us to examine how matching performance changes with the composition of available evidence. Appendix~\ref{apdx:mixed} provides the complete sampling procedure.

\subsection{Matching Methods}
\label{sec:retrieval_methods}

We compare three approaches for matching a query against the candidate MUD profiles.

\textbf{Exact ACE matching.}
The first approach uses literal ACE overlap. We use Jaccard similarity, defined as the size of the ACE intersection divided by the size of the ACE union, and exact ACE hit count, defined as the number of query ACEs found in the candidate profile. 
Exact overlap is the basic matching mechanism used in existing MUD-based monitoring and policy-verification approaches~\cite{tdsc2022,MUDfirewall2024}. We therefore use exact matching as the natural MUD-based baseline for assessing whether semantic similarity provides additional evidence when literal overlap is limited.

\textbf{Aggregated semantic matching.} As a device-level semantic baseline, we mean-pool the whitened ACE embeddings of each candidate profile into a single normalized signature. The query ACE embeddings are aggregated in the same way, and query-to-profile similarity is computed using cosine similarity between the resulting signatures. This baseline tests how much identification evidence remains when individual ACE representations are collapsed into a single device-level vector. Construction details are given in Appendix~\ref{apx:sec:meanpool}.

\textbf{ACE-level semantic matching.}
The third approach retains individual ACE embeddings and applies MaxSim, following the late-interaction principle used by ColBERT~\cite{ColBERT2020}.
Given a query ACE set $Q$ and candidate profile $R_i$, we compute:
\[
    s(Q, R_i) = \frac{1}{|Q|}\sum_{q\in Q}\max_{a\in R_i}\cos(e_q,e_a),
\]
where $e_q$ and $e_a$ denote whitened ACE embeddings. 
For each query ACE, MaxSim finds its most similar ACE in the candidate profile and averages these best-match similarities across the query. The score is asymmetric, meaning that each query ACE is matched against the candidate profile, but candidate ACEs are not required to match the query. This is appropriate because a query may contain only a subset of the behavior represented by the reference profile.

\begin{wraptable}{r}{0.65\columnwidth}
	\centering
	\vspace{-6mm}
	\caption{Identification under controlled zero-overlap conditions. $\Delta$ reports the Top-1 change in percentage points from raw to whitened embeddings for the corresponding semantic method.}
	\vspace{-2mm}
	\resizebox{\linewidth}{!}{%
		\begin{tabular}{lrrrrrr}
			\toprule
			\textbf{Eval Variant} & \textbf{$N$} & \textbf{Jaccard} & \multicolumn{2}{c}{\textbf{Mean Pool}} & \multicolumn{2}{c}{\textbf{MaxSim}} \\
			\cmidrule(lr){4-5} \cmidrule(lr){6-7}
			& & & \textbf{Top-1} & \textbf{$\Delta$} & \textbf{Top-1} & \textbf{$\Delta$} \\
			\midrule
			Single unseen ACE & 1{,}023 & 0.0000 & 0.6393 & +14.7 & 0.6549 & +9.5 \\
			Unseen ACE family & 103 & 0.0000 & 0.8447 & +13.6 & 0.8447 & +4.9 \\
			Unseen ACE set & 240 & 0.0000 & 0.7792 & +25.4 & 0.7958 & +25.0 \\
			\midrule
			Endpoint (all) & 280 & 0.0000 & 0.9393 & +38.2 & 0.9324 & $-0.2$ \\
			Endpoint (domain-rich) & 140 & 0.0000 & 0.9643 & +15.0 & 0.9857 & $-0.2$ \\
			\bottomrule
		\end{tabular}
	}
	\vspace{-3mm}
	\label{tab:controlled_zero_overlap}
\end{wraptable}

\subsection{Results}
\label{sec:device_identification_results}

We report Top-1 accuracy and mean reciprocal rank (MRR). Top-1 measures how often the correct profile is ranked first, while MRR captures how highly the correct profile is ranked across queries. When multiple candidates receive the same positive score, we assign fractional Top-1 credit according to the number of candidates tied for first and compute MRR as the expected reciprocal rank across the tied positions. If all candidate scores are zero, the matcher abstains and receives zero Top-1 and MRR credit.

\paragraph{Unseen ACEs and endpoint perturbation.}
Table~\ref{tab:controlled_zero_overlap} reports the unseen-ACE and endpoint-perturbation conditions. Both conditions are zero-overlap by construction. Unseen query ACEs are removed from the candidate repository, and endpoint perturbations are generated so that they do not reproduce any ACE in the reference corpus.

Exact matching has no identification signal in these conditions because no query ACE exactly matches the candidate repository; therefore, it abstains on every query. These conditions test whether semantic similarity still provides evidence about device identity when literal correspondence is absent.
For single unseen ACEs, MaxSim achieves 0.6549 Top-1 accuracy.
For queries containing multiple unseen ACEs, MaxSim reaches 0.8447 Top-1 for same-family queries and 0.7958 for ACEs drawn from distinct families. Mean pooling performs similarly, reaching 0.6393, 0.8447, and 0.7792 across the single-ACE, same-family, and distinct-family settings. Under endpoint perturbation, MaxSim reaches 0.9324 Top-1 accuracy across all devices and 0.9857 when query sources are restricted to the 14 domain-rich devices; mean pooling reaches 0.9393 and 0.9643, respectively.
The $\Delta$ columns show that whitening substantially improves mean-pooled retrieval, consistent with the reduced cosine concentration observed in \S\ref{sec:whitening}. Whitening has a less consistent effect on MaxSim, ranging from substantial gains in the unseen-ACE conditions to negligible changes under endpoint perturbation.

\begin{wraptable}{r}{0.65\columnwidth}
	\vspace{-4mm}
	\caption{Identification under mixed partial observation over $5,772$ controlled queries.} 
	\vspace{-3mm}
	\label{tab:partial_runtime_drift}
	\centering
	\small
	\resizebox{\linewidth}{!}{%
\begin{tabular}{lcc||cccc}
	\toprule
	\textbf{Method} & \textbf{Top-1} & \textbf{MRR} & \textbf{0 hits} & \textbf{1--2 hits} & \textbf{3--5 hits} & \textbf{$>5$ hits} \\
	\midrule
	Jaccard         & 0.6537 & 0.7195 & 0.0000 & 0.3584 & 0.5908 & 0.8806 \\
	Exact ACE-hit count & 0.7248 & 0.7716 & 0.0000 & 0.3801 & 0.6939 & 0.9623 \\
	MaxSim          & 0.9190 & 0.9348 & 0.7093 & 0.7817 & 0.9287 & 0.9945 \\
	Mean Pool       & 0.9425 & 0.9585 & 0.9075 & 0.8416 & 0.9324 & 0.9978 \\
	\bottomrule
\end{tabular}
}
\vspace{-2mm}
\end{wraptable}

The successful unseen-ACE cases illustrate how semantic matching can recover related communication behavior after exact correspondence is removed. In a single-unseen-ACE episode, the held-out ACE for the HP Printer device was a local IPv4 UDP multicast to {\myverbS{224.0.0.252/32}} on port 5355. This ACE was removed from every candidate profile before scoring, so exact matching had zero overlap. MaxSim nevertheless ranked this device first because the retained HP Printer profile contained a corresponding local IPv6 multicast behavior on the same UDP service.
A multi-ACE unseen-set episode shows the same effect across distinct behaviors. The held-out HP Printer query combined cloud communication ({\myverb{xmpp009.hpeprint.com:80}}), local service discovery on port 5353, and DHCPv4 behavior. After all three exact ACEs were removed from every reference profile, MaxSim still ranked the HP Printer profile first by matching the query to retained ACEs with similar protocol-port behavior and communication roles (\ie HP ePrint traffic on port 443, IPv6 local discovery, and DHCPv6).

\paragraph{Mixed partial observation.}
Table~\ref{tab:partial_runtime_drift} reports the mixed partial-observation condition, where queries contain different combinations of exact, endpoint-perturbed, and unseen ACEs. The Top-1 and MRR columns report overall identification performance across all queries. The remaining columns group queries by the number of exact ACE matches retained against the correct reference profile.

Exact ACE matching performs well when a query retains multiple exact matches to the source profile. In the $>5$ hit bin, Jaccard and exact ACE-hit count achieve 0.8806 and 0.9623 Top-1 accuracy, respectively. Their performance decreases as the exact overlap becomes sparse. With zero exact hits, both methods abstain on every query. In contrast, MaxSim and mean pooling retain Top-1 accuracies of 0.7093 and 0.9075, respectively.
Both semantic methods remain effective as exact overlap decreases. Mean pooling performs best in this mixed condition, achieving 0.9425 Top-1 compared with 0.9190 for MaxSim, and retains 0.9075 Top-1 even in the zero-hit bin. Therefore, retaining individual ACE representations does not mean that MaxSim is always superior to mean pooling. Their relative performance depends on the evidence available in the query.
These controlled experiments characterize matching behavior as the amount and composition of exact ACE evidence are manipulated. We next evaluate the same matching approaches using observations reconstructed from real IoT traffic traces.


\section{Real Traffic Evaluation}\label{sec:runtime_flows}

The previous section characterized exact and semantic matching as the amount and composition of ACE overlap were systematically manipulated.
We now evaluate the same matching approaches using observations reconstructed from real IoT traffic traces, where the available behavioral evidence is determined by the device's observed communication instead of constructing controlled queries.
In this evaluation, the reference MUD profiles remain unchanged, and observed flows are converted into ACE-like behavioral primitives and processed in their original temporal order.
We examine how exact matching, mean pooling, and MaxSim behave as evidence accumulates and within short traffic windows.

{\subsection{Runtime Flow Reconstruction}}

We use public traffic traces from~\cite{dryadDataset} and the reference MUD profiles described in \S\ref{sec:representation}~\cite{tdsc2022}. We exclude the Withings Smart Scale trace because no corresponding MUD profile is available. The Chromecast profile has no corresponding traffic trace. This leaves 26 device traces, each identified against all 28 candidate MUD profiles.
For each trace, we retain IP flows containing transport protocol and port information. Non-IP traffic and flows without transport-layer ports, such as ARP, are excluded. DHCP, multicast discovery, and other local transport communications are retained when they satisfy these criteria. We use up to the first 50,000 flows from each trace in arrival order, yielding 810,490 flows in total. 

Each retained flow is converted into an ACE-like behavioral primitive using the same protocol, endpoint, direction, and port fields represented by the reference ACEs. Flows are oriented as ingress or egress relative to the device, and remote IP addresses are mapped to hostnames whenever the corresponding DNS information is available in the  trace. When no hostname can be resolved, we retain the remote IP address using a /32 prefix for IPv4 or /128 for IPv6. This process resolves 3,286 of 3,487 unique device-remote IP pairs (94.2\%). By flow volume, 284,851 of 810,490 flow instances (35.1\%) are hostname-resolved. The remainder includes repeated communication with local, broadcast, multicast, gateway, or otherwise unresolved endpoints.

\begin{wrapfigure}{r}{0.60\textwidth}
	\vspace{-4mm}
	\centering
	{\includegraphics[width=\linewidth]{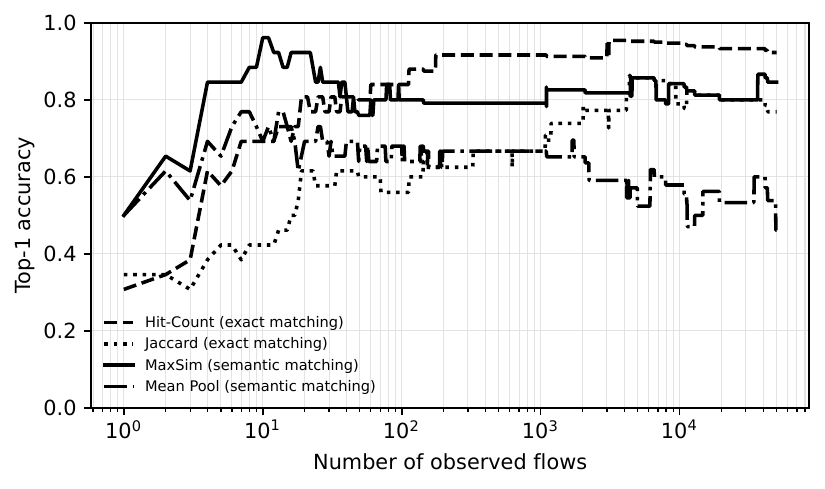}}
	\vspace{-7mm}
	\caption{Top-1 accuracy during cumulative real-traffic identification as the count of observed flows increases.}
	\vspace{-4mm}
	\label{fig:runtime_accuracy_convergence}
\end{wrapfigure}

We apply the same matching strategies as in \S\ref{sec:retrieval_methods}. Exact matching uses normalized ACE hit count and Jaccard similarity, while semantic matching uses either a mean-pooled signature or MaxSim over individual ACEs. For hit count, we normalize the number of exact matches by the number of observed primitives, including repeated observations; we refer to this value as the exact-overlap score. Repeated flow observations are retained for hit count, Mean Pool, and MaxSim, allowing recurring behaviors to contribute repeatedly to the accumulated evidence; Jaccard operates over unique ACE observations. All semantic methods use the same BGE-M3 embeddings and the whitening transformation from the reference corpus, as described in \S\ref{sec:whitening}.

\subsection{Identification as Evidence Accumulates}

We first evaluate identification as flows accumulate in their original temporal order. At observation point $k$, the first $k$ retained flows of each device trace form the observed behavior. Fig.~\ref{fig:runtime_accuracy_convergence} reports Top-1 accuracy as the amount of observed traffic increases.

The matching approaches exhibit different behavior as evidence accumulates.
After a single flow, MaxSim and Mean Pool each identify $13/26$ traces, while exact hit-count matching identifies $8/26$. 
After observing 10 flows, MaxSim reaches $25/26$, compared with $18/26$ for both hit-count and Mean Pool.
MaxSim, therefore, extracts useful identification evidence particularly quickly from the early observations.
As additional traffic accumulates, repeated exact matches increasingly strengthen hit-count matching, which eventually identifies $25/26$ traces at the end of the observation period. MaxSim remains correct for $24/26$, whereas Mean Pool finishes at $17/26$. Jaccard, which operates over unique ACE observations rather than repeated evidence, remains weaker than the hit count for much of the trace.

Mean Pool and MaxSim behave differently as more traffic is observed.
Although both identify 13/26 traces after the first flow, their performance diverges as additional observations accumulate. Mean Pool compresses all observed primitives into a single signature, whereas MaxSim retains them individually for late interaction with each candidate profile. We examine this distinction further using bounded traffic windows in \S\ref{sec:runtime_windows}.

\begin{wrapfigure}{r}{0.60\textwidth}
	\centering
	{\includegraphics[width=\linewidth]{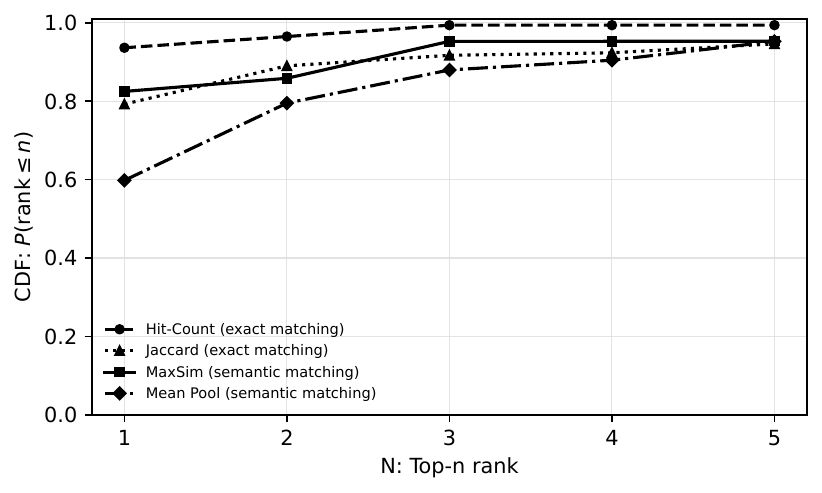}}
    \vspace{-6mm}
	\caption{Distribution of the correct-device rank during cumulative real-traffic identification over the first 10,000 observed flows of each trace.}
    \vspace{-5mm}
	\label{fig:runtime_rank_distribution}
\end{wrapfigure}

Top-1 accuracy does not show whether an incorrect prediction nevertheless places the correct device among a small set of highly ranked candidates. We therefore examine the distribution of the rank of the correct device across the first 10,000 observed flows for each trace. This per-trace cap prevents devices with longer traces from contributing disproportionately more rank observations to the distribution. Fig.~\ref{fig:runtime_rank_distribution} reports the resulting cumulative rank distribution.

Exact hit count most frequently places the correct profile first. Among the semantic methods, MaxSim provides substantially stronger Top-1 ranking than Mean Pool ($0.825$ versus $0.598$). The gap narrows at larger rank cutoffs: MaxSim places the correct profile within the top three for $0.953$ of observations compared with $0.880$ for Mean Pool, while both reach $0.953$ within the top five. Therefore, mean pooling often retains the correct device among a small candidate set but provides weaker discrimination at the top of the ranking.

\subsection{Identification from Short Traffic Windows}
\label{sec:runtime_windows}

Cumulative evaluation favors methods that can accumulate evidence over long traces and depends on the starting point of each capture. We therefore evaluate identification using short, fixed-size observations drawn from different points in each trace, rather than allowing evidence to accumulate from the start of capture. 
For each device, we sample up to 500 disjoint windows containing 50 consecutive flows\footnote{The \texttt{Blipcare BP Meter} trace contained fewer than 50 eligible flows and is excluded from this experiment.}, yielding 9023 runtime windows across 25 devices. 
Each window is treated as an independent identification query against the same 28 reference MUD profiles. 

Table~\ref{tab:runtime_start_windows} reports identification over all windows.
Across 9023 windows, MaxSim identifies 6823 (75.6\%) correctly, compared with 5707 (63.3\%)  for hit count and 5024 (55.7\%) for Mean Pool. Thus, MaxSim achieves the highest Top-1 accuracy when identification uses only 50 consecutive flows.

The difference becomes more pronounced in windows with limited exact evidence (second--fourth rows in Table~\ref{tab:runtime_start_windows}).
When the exact-overlap score is below 0.50, MaxSim identifies 2099 of 3287 windows (63.9\%), compared to 893 (27.2\%) for exact matching and 1034 (31.5\%) for Mean Pool. For scores below 0.10, MaxSim remains correct for 702 of 1841 windows (38.1\%), whereas exact matching and Mean Pool identify only 40 (2.2\%) and 49 (2.7\%), respectively. Finally, among the 1152 windows from seven devices with no exact overlap, MaxSim identifies 457 (39.7\%), compared with 13 (1.1\%) for Mean Pool; exact matching has no positive evidence in this condition.

\begin{table}[t!]
    \caption{Identification over 9,023 randomly sampled, non-overlapping
    50-flow windows. The last three columns report the number and percentage
    of windows correctly identified at Top-1.}
    \label{tab:runtime_start_windows}
    \vspace{-2mm}
    \centering
    \small
    \renewcommand{\arraystretch}{1.05}
    \resizebox{0.985\linewidth}{!}{
        \begin{tabular}{lrrrrr}
            \toprule
            \textbf{Window set} &
            \textbf{Windows} &
            \textbf{Devices} &
            \multicolumn{3}{c}{\textbf{Top-1 correct}} \\
            \cmidrule(lr){4-6}
            & & &
            \textbf{Exact} &
            \textbf{Mean Pool} &
            \textbf{MaxSim} \\
            \midrule
            All windows
            & 9{,}023 & 25
            & 5{,}707 (63.3\%)
            & 5{,}024 (55.7\%)
            & 6{,}823 (75.6\%) \\
    
            Exact-overlap score $<0.50$
            & 3{,}287 & 14
            & 893 (27.2\%)
            & 1{,}034 (31.5\%)
            & 2{,}099 (63.9\%) \\
    
            Exact-overlap score $<0.10$
            & 1{,}841 & 8
            & 40 (2.2\%)
            & 49 (2.7\%)
            & 702 (38.1\%) \\
    
            Exact-overlap score $=0$
            & 1{,}152 & 7
            & 0 (0.0\%)
            & 13 (1.1\%)
            & 457 (39.7\%) \\
            \bottomrule
        \end{tabular}
    }
\end{table}

Individual windows illustrate the type of evidence that semantic matching retains when literal overlap is weak.
For example, in one \texttt{HP Printer} window, multiple flows contact ``{\myverbS{xmpp006.hpeprint.com:\\5222}}'', whereas the reference profile contains ``{\myverbS{xmpp009.hpeprint.com:5222}}''. MaxSim nevertheless ranks \texttt{HP Printer} first by associating the related ePrint communication behaviors. Similarly, a \texttt{Triby Speaker} window contains ``{\myverbS{sip.invoxia.com:5228}}'' while its reference profile contains ``{\myverbS{sip.aws.invoxia.io:5228}}''; MaxSim preserves useful similarity through their shared protocol, port, and endpoint semantics.

We note that the relative performance of the two semantic strategies differs between the controlled and real-traffic evaluations. Under the controlled mixed-partial-observation experiment, Mean Pool achieves 0.9425 Top-1 compared with 0.9190 for MaxSim, whereas on real 50-flow windows, MaxSim reaches 0.756 compared with 0.557 for Mean Pool. This reversal is consistent with the greater heterogeneity of real observations, which include repeated communication, shared services, local communication, unresolved endpoints, and behaviors not represented in the reference profile. Mean Pool combines these observations into a single signature, whereas MaxSim retains individual primitives during matching. Performance under controlled overlap degradation does not fully predict behavior on real traffic. Preserving individual primitives provides substantially stronger identification on the heterogeneous runtime observations evaluated here.

\section{Related Work}

\textbf{IoT device identification from network traffic.} Prior work has shown that IoT devices expose recognizable communication patterns that support device fingerprinting and classification~\cite{sivanathan2019,20IoTJ1Class,haystack2020IMC,25tnse}. Existing approaches typically learn signatures from packets or flows using statistical traffic features, protocol sequences, or endpoint information~\cite{IPbased2018,22ipfixJIoT}. More recent work applies representation learning to encrypted traffic using transformer architectures and self-supervised pre-training \cite{lin2022etbert,zhao2023yATC,wang2024netmamba,zhou2024netflowgen}. These methods use packet-level \cite{pasquini2025} or flow-level \cite{25tnse} representations.
Our work studies a complementary abstraction. Instead, we represent communication behavior using Access Control Entries (ACEs), which encode protocol, endpoint, direction, and port semantics. ACEs provide a standardized behavioral abstraction that can be reconstructed from runtime traffic and compared against reference MUD profiles. Our objective is not to replace packet- or flow-based classifiers, but to investigate MUD-derived behavioral primitives as an additional identification signal.

\textbf{MUD-based monitoring and runtime verification.} The Manufacturer Usage Description (MUD) standard~\cite{rfc8520} provides a machine-readable specification of intended IoT communication behavior. 
Prior work has operationalized MUD for runtime monitoring, policy verification, and enforcement by comparing observed communication against MUD-defined rules~\cite{tdsc2022,MUDfirewall2024,ACSAC2022}.
These approaches primarily treat ACEs as policy rules matched literally against observed behavior. Our work instead asks whether ACEs can also serve as behavioral primitives for device identification when literal correspondence is limited. We characterize exact and semantic matching under controlled changes in ACE overlap and then evaluate the same representations using ACE-like observations reconstructed from real traffic.

\textbf{Semantic encoders and late interaction retrieval.}
Embedding models such as BERT and SBERT are widely used for semantic retrieval~\cite{ddevlin2019bert,reimers2019sbert}, while recent encoders such as BGE-M3 support longer structured texts and cosine-based retrieval objectives~\cite{chen2024bgem3}. Whitening can reduce embedding anisotropy and cosine concentration~\cite{li2020bertflow,su2021whitening,beltiukov2025demystifying}. We apply these techniques to ACE-level behavioral representations, using reference-derived whitening to calibrate semantic similarity. Our ACE-level matching method further adopts the late-interaction principle of ColBERT~\cite{ColBERT2020}, retaining individual behavioral primitives during matching rather than first collapsing them into a single aggregated signature.

\section{Conclusion}

This paper investigated MUD-based IoT device identification using individual ACEs as behavioral primitives for exact and semantic matching. We first compared different ways of representing 28 MUD profiles and found that keeping ACEs separate preserves variation among communication rules that is lost when the whole profile is represented by a single vector. We then used controlled queries to show that semantic similarity retains identification evidence when the literal ACE overlap is reduced through unseen behavior, endpoint perturbation, and partial observation.
Finally, our evaluation on more than 800,000 real IoT flows showed that exact and semantic matching provide complementary signals. Exact matching becomes strong as repeated literal evidence accumulates, while semantic matching of individual ACEs performs better  during early observations and when identification is based on short traffic windows  with limited exact overlap. Mean pooling performs strongly on controlled queries but degrades substantially on heterogeneous real observations, highlighting the value of retaining individual primitives during matching. Semantic ACE matching complements exact MUD matching and existing methods that identify devices from network traffic. It is particularly useful when reference MUD profiles are available, but literal correspondence with the observed traffic is limited.

\bibliographystyle{ACM-Reference-Format}
\bibliography{embeddingMUD}

\appendix
\section{Embedding Calibration Details}
\label{apdx:whit-detl}

\subsection{Whitening and Normalization}\label{apx:sec:white}

All semantic matching experiments use whitened ACE embeddings. Whitening parameters are estimated exclusively from the reference ACE corpus; controlled queries and runtime observations are not used to estimate the transformation. Let $\mu$ and $\Sigma$ denote the mean and covariance of the reference embeddings, with eigendecomposition:
\[
    \Sigma = U \Lambda U^\top
\]
For a raw embedding $x \in \mathbb{R}^{1024}$, we retain the $k=256$ principal components associated with the largest eigenvalues and compute

\[
    \tilde{x}
    =
    \Lambda_k^{-1/2}
    U_k^\top
    (x-\mu),
\]

The resulting vector is $L_2$-normalized before similarity computation. The same reference-derived transformation is applied unchanged to candidate-profile ACEs and query ACEs. Estimating the transformation only from the reference corpus prevents information from evaluation queries being used in the embedding calibration.

\subsection{Mean-Pooled Signatures}\label{apx:sec:meanpool}

The aggregated-signature baseline represents an ACE set $A$ using a single vector. Given the whitened and $L_2$-normalized embedding $\tilde{x}_a$ of ACE $a$, we first compute

\[
    \mu(A)
    =
    \frac{1}{|A|}
    \sum_{a\in A}\tilde{x}_a
\]

and then normalize the pooled representation:

\[
    v(A)
    =
    \frac{\mu(A)}
         {\|\mu(A)\|_2}.
\]

Candidate signatures are constructed from all ACEs in each reference MUD profile, while query signatures are constructed from the ACEs contained in the corresponding query. The similarity between a query and candidate profile is computed using cosine similarity between their normalized signatures.

\section{Controlled Query Construction Details}\label{apdx:eval-proto}

This appendix provides reproducibility details for the controlled query construction described in \S\ref{sec:device_identification}. 

\subsection{Unseen ACEs}\label{apdx:unseen}

The unseen-ACEs condition constructs queries in which the selected ACEs have no exact counterpart in the candidate repository. We evaluate three variants.
\textbf{Single unseen ACE:}
Each of the 1,023 $(\text{device},\text{ACE})$ pairs 
defines one episode; no device contains duplicate compact ACE text. 
The query contains that ACE alone; its text is removed (before scoring) from every candidate profile in which it occurs. 
This yields 1,023  zero-overlap episodes.
\textbf{ACE-family construction:}
We cluster 710 unique ACE texts using whitened embeddings. 
Two ACEs are connected when each appears among the five nearest neighbors with cosine similarity of at least 0.75; connected components define ACE families. 
This produces 590 families, 
including 101 non-singleton families. 
\textbf{Unseen ACE family:}
Singleton families are excluded because they are already covered by the single-ACE condition. We group ACE instances by $(\text{device},\text{family})$ and retain groups where a device contains at least two distinct ACEs from the same family. The query contains the eligible family members present in the source device, while the queried ACE texts are removed from every candidate profile before scoring. This produces 103 episodes from 99 families across 18 devices; queries contain between two and four ACEs. 
\textbf{Unseen ACE set:}
For each eligible device, we sample three distinct families and one ACE uniformly from each family, producing a three-ACE query. 
The selected ACEs are removed from the candidate repository before scoring.
Devices with fewer than 12 ACEs are excluded; 24 devices remain. Ten seeded queries per  device yield $24\times10=240$ episodes.

\subsection{Endpoint Perturbation}\label{apdx:perturb}
For each selected domain-name ACE, we  determine the applicable perturbations and randomly select one (uniformly). 
\textbf{Numeric modification} either increments an existing number (\eg {\myverbS{time1}} $\rightarrow$ {\myverbS{time2}}) or appends a digit from $\{2,3,4\}$. \textbf{Regional insertion} adds a token from $\{${\myverbS{us}}, {\myverbS{eu}}, {\myverbS{ap}}, {\myverbS{na}}, {\myverbS{emea}}$\}$ either within or as a new hostname label. \textbf{Domain-suffix substitution} replaces the existing suffix using a predefined source-dependent mapping. The complete perturbation rules and substitution map are provided with the released code \cite{gitHub2026}.
Each resulting ACE is checked against all ACE texts in the reference corpus; collisions are regenerated for up to eight attempts. No evaluated mutation exhausted this limit. Protocol, direction, and port remain unchanged. 




\textbf{Episode construction.}
For each device, we 
sample 10\% of its perturbable domain-name ACEs (minimum one and maximum three), perturb the selected endpoints, and use only those ACEs as the query. We generate ten variants per device. 
\textit{Endpoint (all)} uses all 28 source devices, yielding $28\times10=280$ episodes; \textit{Endpoint (domain-rich)} uses the 14 devices with at least ten perturbable domain-name ACEs, yielding $14\times10=140$ episodes. Every query is scored against the same 28 candidate profiles, and no generated episode was discarded.

\subsection{Mixed Partial Observation}\label{apdx:mixed}

Mixed queries vary the retained fraction $r\in\{0.10,0.25,0.50\}$, endpoint-perturbation fraction $p\in\{0.00,0.25,0.50,1.00\}$, and unseen-ACE setting $k\in\{0,1\}$. For each source device, we first sample a fraction $r$ of its reference ACEs. We then perturb a fraction $p$ of the sampled ACEs with eligible domain names using Appendix~\ref{apdx:perturb}. When $k=1$, one sampled ACE is selected as unseen, and its original text is removed from the source device's reference profile for that query. The unseen ACE is selected after endpoint perturbation and may therefore also be endpoint-perturbed. Unlike the strict unseen-ACE experiments in Appendix~\ref{apdx:unseen}, this removal applies only to the source profile.

The full grid contains $3\times4\times2\times28\times10=6720$ possible queries. We exclude 240 cases where making one ACE unseen would leave no known ACE in the query and 708 cases where endpoint perturbation is requested, but the sampled query contains no perturbable hostname ACE. No query is excluded because of an empty sample or failed endpoint mutation. The resulting evaluation contains $6720-948=5772$ controlled queries.

\end{document}